\documentclass[twocolumn]{revtex4}
\usepackage{graphicx}

\begin{document}
\title{Emergence and decay of turbulence in stirred atomic Bose-Einstein condensates}
\author{N.G. Parker and C.S. Adams}
\affiliation{Department of Physics, University of Durham, South
Road, Durham, DH1 3LE, United Kingdom}

\begin{abstract}
We show that a `weak' elliptical deformation of an atomic
Bose-Einstein condensate rotating at close to the quadrupole
instability frequency leads to turbulence with a Kolmogorov energy
spectrum. The turbulent state is produced by energy transfer to
condensate fragments that are ejected by the quadrupole
instability. This energy transfer is driven by breaking the
two-fold rotational symmetry of the condensate. Subsequently,
vortex-sound interactions damp the turbulent state leading to the
crystalization of a vortex lattice.
\end{abstract}

\maketitle

Two-dimensional (2D) turbulence has been explored in diverse areas
such as soap films \cite{soapfilms}, magnetohydroynamics
\cite{cardoso}, and meteorology (see \cite{smith} and references
therein), and can often display additional features not present in
3D \cite{klb}. However, the complexities of real fluids mean that
the theoretical predictions are often at odds with observed
spectra. In this regard there may be some advantages in studying
superfluids where the absence of viscosity and the quantization of
vorticity can simplify the theoretical picture. For example,
superfluid turbulence in liquid Helium \cite{barenghi} is found to
exhibit analogous features to classical turbulence, in particular
a Kolomorgorov energy spectrum \cite{sfturbulence}. Even more
amenable to theoretical description are atomic Bose-Einstein
condensates (BECs) \cite{BEC}. In addition, atomic BECs allow the
flexibility of studying the transition between 2D and 3D
turbulence.

Recent experiments on atomic BECs have generated vortex lattices
by thermodynamically condensing a rotating thermal cloud
\cite{haljan} and  mechanical rotation of the condensate in an
anharmonic trap \cite{madison2000,abo-shaeer2001,hodby}. In the
latter case, a quadrupolar collective mode of the condensate is
excited.  A dynamical instability \cite{madison2001,sinha} leads
to the nucleation of vortices, which subsequently crystallise into
a lattice configuration. The time-scale for vortex lattice
formation has been shown to be insensitive to temperature
\cite{abo-shaeer2002,hodby}, suggesting that the process is a
purely dynamical effect. The formation of the lattice has been
simulated using the time-dependent Gross-Pitaveskii equation (GPE)
in 2D with the inclusion of damping effects
\cite{tsubota,penckwitt}, and in 3D \cite{lobo}. However,
questions remain over the dynamics involved, for example, how the
dynamical instability seeds vortex nucleation, what is the damping
mechanism leading to the crystallization of the lattice, and the
role of dimensionality in the process.

In this paper, we present evidence that 2D turbulence is a key
feature of current experiments on vortex lattice formation in
atomic BECs. In particular we show that the route to lattice
formation can be divided into four distinct stages, as illustrated
in Fig.~1. These stages are:

- {\it Fragmentation:} The quadrupolar mode breaks down,
    ejecting energetic atoms to form an outer cloud.

- {\it Symmetry-breaking:} The two-fold rotational symmetry of the
    system is broken in a macroscopic manner, allowing the rotation to
    couple to additional modes, thereby injecting energy rapidly into
    the system.

 - {\it Turbulence:}  A turbulent cloud containing vortices
    and high energy density fluctuations (sound field) is formed, with a Kolmogorov energy spectrum.

 - {\it Crystallisation:}  The loss of energy at short length-scales coupled with the vortex-sound interactions \cite{vortexsound1,vortexsound2} allow the system to relax into an
ordered lattice.

\begin{figure}
\includegraphics[width=4.7cm,angle=-90]{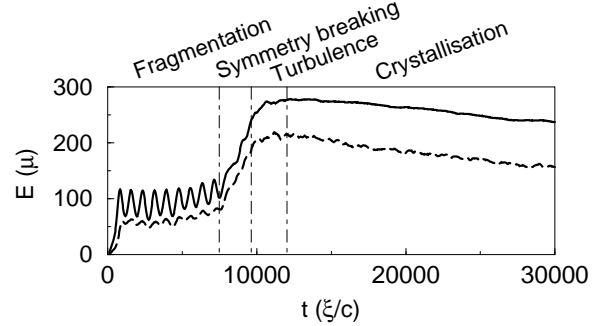}
\caption{The condensate energy (black line) and energy of the
outer cloud (dashed line), defined as the region with $n<0.2n_0$,
as a function of time, with a trap rotation frequency
$\Omega=0.77\omega_r$. The four distinct stages of the evolution
(see text) are indicated. }
\end{figure}

Our analysis is based on the classical field method
 whereby the GPE is
used to describe both the condensate mean-field $\psi$ and
fluctuations \cite{norrie05}. Fluctuations in the initial state
speed-up the evolution through the four stages but do not change
the qualitative behaviour. For illustrative purposes we use an
initial state without excitations.  The rotating trap strongly
polarizes the dynamics along the $z$-axis, such that 2D dynamics
in the rotating plane dominate the system. Indeed we have verified
that the same results are obtained in the flattened, quasi-2D
geometry by solving the 3D GPE. We therefore proceed in the 2D
limit by solving the computationally advantageous 2D GPE, as in
our previous work \cite{vortexsound1}. We have performed
simulations for a range of condensate sizes, and various grid and
box sizes, and find the same qualitative results throughout.  We
characterise the state of the system in terms of the energy, which
is calculated using the time-independent GPE energy functional,
\begin{eqnarray}
E=\int\left[-\frac{\hbar^2}{2m}|\nabla
\psi|^2+V_{\rm{ext}}\psi^2+\frac{g}{2}|\psi|^4 \right]{\rm d}x
{\rm d}y,
\end{eqnarray}
and subtract the corresponding energy of the initial state. Here
$V_{\rm ext}$ is the external potential, $m$ is the atomic mass,
and $g=4\pi \hbar^2 Na/m$, where $N$ is the number of atoms and
$a$ is the {\it s}-wave scattering length.

To apply a rotation we use a `weakly' elliptical potential similar
to the experiments \cite{madison2000,hodby} (the stirring
potential used in Ref. \cite{abo-shaeer2001} is strongly
anharmonic). The potential can be expressed as,
\begin{equation}
V_{\rm ext}(x,y,t)=\frac{1}{2}m\omega_r^2 \left[(x^2+y^2)
+\varepsilon \left(X^2-Y^2\right)\right].
\end{equation}
The first term represents the static harmonic trap with transverse
 frequency $\omega_r$. In the second
term, the parameter $\varepsilon$ characterises the trap
anisotropy in the coordinate system $(X, Y)$ which is aligned with
the static $(x,y)$ coordinates at $t=0$ and subsequently rotates
at angular frequency $\Omega$. We employ a trap deformation of
$\varepsilon=0.025$ \cite{madison2000}. Whereas previous
theoretical investigations have considered the rotating frame
\cite{tsubota,penckwitt,lobo}, we perform simulations in the
laboratory frame.

Our units of length, time and energy are the healing length
$\xi=\hbar/\sqrt{m \mu}$, $(\xi/c)$, and the chemical potential
$\mu-n_0g$. Here $c=\sqrt{\mu/m}$ is the Bogoliubov speed of
sound, $\mu=n_0g$ is the chemical potential and $n_0$ is the peak
condensate density. Using typical parameters for a $^{87}$Rb
condensate, the units of distance and time correspond to $\xi \sim
0.3~{\rm \mu m}$ and $(\xi/c)\sim 10^{-4}{\rm s}$, respectively.

{\it Fragmentation:} At $t=0$, the rotation is turned on. The
rotating trap couples energy into the condensate by exciting a
quadrupolar shape oscillation, while the axes of the quadrupole
rotate at the trap rotation frequency $\Omega$. This excitation
mimics rotation yet the system remains irrotational.  In the
radially-symmetric system, the quadrupole mode is predicted to
have a resonant frequency at $\Omega=\omega_r/\sqrt{2}$
\cite{stringari}, but this is shifted slightly by the ellipticity.
Away from the resonant frequency, the quadrupole oscillations have
reduced amplitude and are stable. The condensate cycles between
the initial circular state and a higher energy elongated state,
but over a complete cycle there is no {\it net} increase in
energy. Such quadrupole oscillations have been observed
experimentally \cite{quad_expts}.

The condensate is predicted to be dynamically unstable at the
quadrupole resonant frequency \cite{sinha}. The instability arises
because the amplitude of the mode becomes so large that the
quadrupolar irrotational flow can no longer be supported. As the
condensate relaxes from the point of maximum elongation, the fluid
cannot adjust sufficiently quickly back towards the centre of
trap. This results in the shedding of fluid from the ends of the
condensate forming low density tails and giving the condensate a
spiral shape (Fig. 2(a)(ii)). The ejected material collapses back
onto the condensate edge, forming phase dislocations with the main
condensate.  This generates surface waves and `ghost' vortices
\cite{tsubota}. An outer, low density ($\sim 0.1n_0$) cloud is
formed (Fig. 2(a)(iii)). After the condensate has shed material
(Fig. 2(b), solid line) its energy no longer decreases back to the
inital value - energy has been transferred irreversibly into the
condensate. We monitor the relative evolution of the outer, low
density cloud and the inner, high density condensate by defining
the outer cloud to be where the density is less than a certain
value, taken here to be $0.2n_0$. The injected energy goes
primarily into the outer cloud (Fig. 2(b), dashed line). Note
that, although the energy of the inner and outer clouds are
comparable, the outer cloud contains only around $10\%$ of the
total atoms, and therefore the average energy per atom is
considerably higher. Subsequently, the inner cloud continues to
undergo quadrupole oscillations and eject small fragments, as
indicated by the energy curves shown in Fig. 2(b). Also, the outer
cloud develops more structure at short length-scales
(Fig.~2(a)(iv)).

\begin{figure}
\includegraphics[width=8cm]{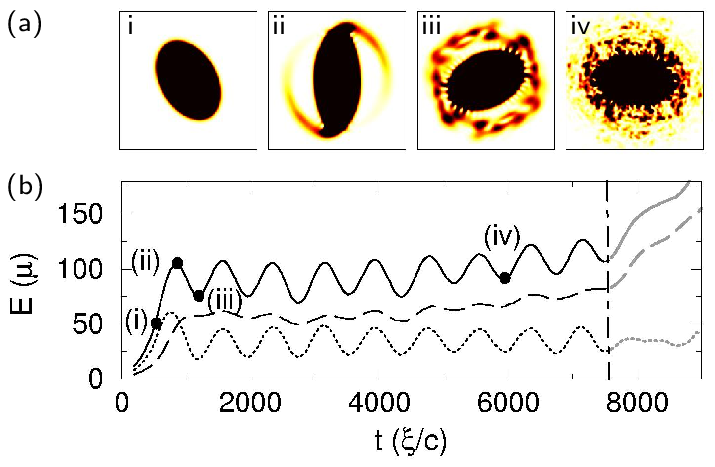}
\caption{Fragmentation: (a) Snapshots of the condensate density in
the range $(0-0.05n_0)$ at times $t\approx~$(i) 400, (ii) 800,
(iii) 1200, and (iv) 6000 $(\xi/c)$. Each plot represents a region
$[-30,30]\xi\times[-30,30]\xi$ (while the numerical box is much
larger). Dark represents high density. (b) Total condensate energy
(solid line), energy of inner cloud (dotted line), and energy of
outer cloud (dashed line). }
\end{figure}

For the parameters employed here, we observe the fragmentation of
the cloud (and ultimately the formation of a vortex lattice) for
rotation frequencies in the range $0.72<\Omega/\omega_r<0.78$,
which is in reasonable agreement with the experimental results of
Madison {\it et al.} \cite{madison2000}. Outside this range the
quadrupole mode is stable, and the width of the unstable region
increases with the trap ellipticity $\varepsilon$, which is
consistent with the experimental observations of Hodby {\it et
al.} \cite{hodby}.

{\it Symmetry breaking:} The condensate and potential have a
two-fold rotational symmetry, which is clearly visible in
Fig.~2(a). In the experiments \cite{madison2000,hodby} this
symmetry is absent. Eventually in the simulations, an asymmetry
grows out of the numerical noise generated by modelling a rotation
using a static square grid. We characterise this asymmetry in
terms of an asymmetry parameter defined as,
\begin{equation}
\sigma=\frac{\int\left[|\psi \left(x,y \right)|^2- |\psi
\left(-x,-y \right)|^2\right]{\rm d}x{\rm d}y}{\int |\psi
\left(x,y \right)|^2 {\rm d}x{\rm d}y}.
\end{equation}
\begin{figure}
\includegraphics[width=7.5cm]{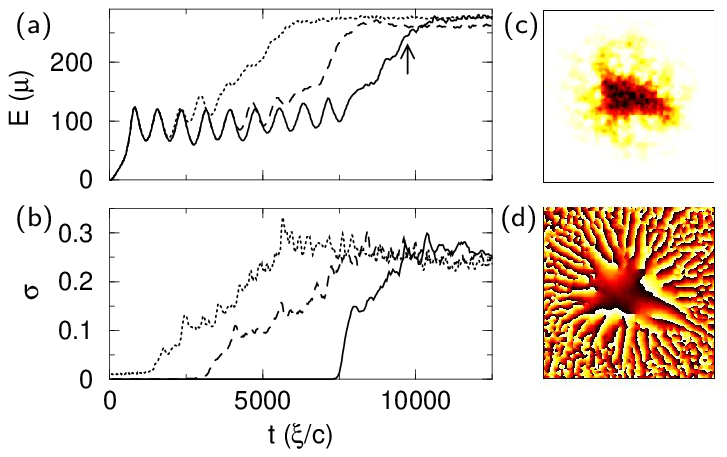}
\caption{Symmetry breaking: Evolution of (a) total condensate
energy and (b) condensate asymmetry parameter for the case of no
trap jitter (solid line), and trap jitters of $\gamma=0.0001$
(dashed line) and $\gamma=0.1$ (dotted). (c)-(d) Snapshots of the
total condensate density and phase, respectively, during the
symmetry breaking stage at $t=9600~(\xi/c)$ in the absence of trap
jitter.}
\end{figure}
This asymmetry parameter (Fig.~3(b), solid line) grows
exponentially over time. When it reaches macroscopic levels
($\sigma \sim 0.1$) we enter a symmetry-breaking phase at
$t\sim8000~(\xi/c)$. From then on, additional modes can be
excited, and energy is rapidly coupled into the system (Fig.~3(a),
solid line). This energy predominantly excites the outer, low
density cloud, as shown in Fig.~1 (dashed line). The condensate
density and phase during this stage are shown in Fig.~3(c)-(d).
The original two-fold rotational symmetry is now clearly broken.
Towards the end of this stage the outer cloud strongly couples to,
and merges with, the inner condensate. Energy and angular momentum
become transferred to the inner cloud, the quadrupole mode finally
breaks down, and vortices become nucleated in the higher density
regions.

In experiments, symmetry-breaking will occur due to, for example,
the thermal cloud, quantum fluctuations, trap imperfections, and
fluctuating fields.  We model the effect by allowing the trap
centre to randomly jump, or jitter, within a region
$[-\gamma,+\gamma]\times[-\gamma,+\gamma]$. Fig. 3(a)-(b) show the
results for trap jitters of $\gamma=0.0001$ (dashed line) and
$0.1$ (dotted line). We observe the same {\it qualitative}
behaviour as in the absence of the jitter, although the
macroscopic symmetry-breaking occurs earlier when jitter is added.
Even for an extremely small jitter ($\gamma=0.0001\xi$, 3 orders
of magnitude smaller than the grid size) the effect is
significant, thereby demonstrating the importance of symmetry.
Symmetry-breaking allows vortices to enter the condensate one by
one, rather than in opposing pairs \cite{tsubota} which reduces
the threshold energy for vortex nucleation.

{\it Turbulence:} Following the injection of energy into the
condensate during the symmetry-breaking stage, a highly excited
and energetic condensate containing randomly-positioned vortices
is formed, as shown in Fig.~4(a)-(b). We analyse this stage of the
evolution by calculating the energy spectrum, shown in Fig. 4(c).
\begin{figure}
\includegraphics[width=8cm]{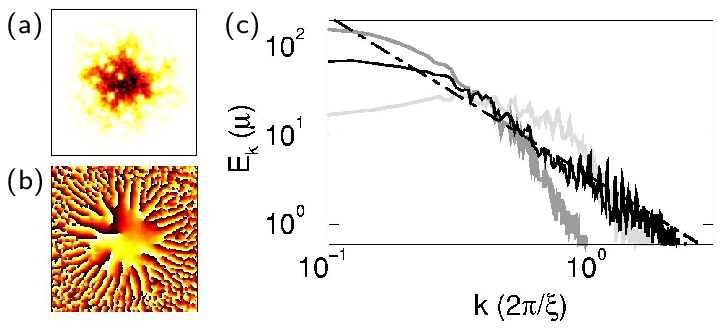}
\caption{Turbulence:  Snapshots of the turbulent (a) condensate
density and (b) phase at $t=11000~(\xi/c)$. The vortices are
characterised by a node in the density and an azimuthal phase
change of $2\pi$. (b) Energy spectrum in {\it k}-space during: (i)
the turbulent stage at $t=11000~(\xi/c)$ (bold line); (ii)
fragmentation at $t=8000~(\xi/c)$ (intermediate grey line); and
crystallisation at $t=20000~(\xi/c)$ (light grey line). The
turbulent Kolmorogov behaviour $E_k\propto k^{-5/3}$ is indicated
(dot-dashed line). }
\end{figure}
During the turbulent phase the system closely follows a Kolmogorov
energy spectrum $E(k)\propto k^{-5/3}$ over a range of {\it
k}-values, as shown in Fig 4(c) (bold line) for $t=11000(\xi/c)$.
Such behaviour is a key signature of classical turbulence and also
occurs in models of superfluid turbulence \cite{sfturbulence}. The
departure from a $k^{-5/3}-$law occurs at an upper bound of
$k\approx2 (2\pi/\xi)$, corresponding to the characteristic
length-scale of vortex-sound interactions \cite{vortexsound2}, and
a lower bound of $k\approx 0.3(2\pi/\xi)$, corresponding to the
size of the condensate. An additional feature of 2D classical
turbulence is that the spectrum is predicted to show $k^{-3}$
behaviour following the $k^{-5/3}$ region \cite{klb}. However, we
observe a $k^{-6}$ dependence in this range. The power spectrum in
the fragmentation stage (Fig. 4(a), intermediate grey) and the
succeeding crystallisation stage (Fig. 4(a), light grey) do not
show a $k^{-5/3}$ behaviour.

{\it Crystallisation:} The turbulent state contains a dense sound
field and vortices. In previous work we have shown how vortices
can both radiate and absorb sound waves in a Bose-Einstein
condensate \cite{vortexsound1,vortexsound2}.  We propose that this
vortex-sound interaction enables the randomly-positioned vortices
in our rotating condensate to crystallise into a lattice.

To demonstrate the transfer of energy from the sound field
(density fluctuations) to the vortices, we divide the energy into
a component due to the vortices $E_{\rm V}$ and a component due to
the sound field $E_{\rm S}$. We approximate $E_{\rm V}$ at a
particular time by numerically generating a similar condensate (by
propagating the GPE in imaginary time) with the same vortex
distribution but without sound. The sound energy is then defined
as $E_{\rm S}=E-E_{\rm V}$. During the fragmentation, symmetry
breaking and turbulent stages, both sound energy (Fig.~5(a),
dotted line) and vortex energy (Fig.~5(a), dashed line) are fed
into the system. At the start of the crystallisation phase the
sound energy is considerably higher than the vortex energy.
However as time progresses the sound energy decreases with this
energy being transferred to the vortices. Figs. 5(b)-(c) shows the
condensate density at the beginning and towards the end of the
crystallisation phase.  We see that the conversion of sound energy
into vortex energy is associated with the ordering of the vortices
from a disordered distribution to a lattice configuration and a
smoothing of the condensate profile (Fig.~5(b)). Furthermore, the
outer cloud shrinks (Fig.5(c)). By the time $t=100000~(\xi/c)$ the
vortex energy is substantially greater than the sound energy.  At
this point we observe a well-ordered vortex lattice.

\begin{figure}[t]
\includegraphics[width=8.5cm,angle=0]{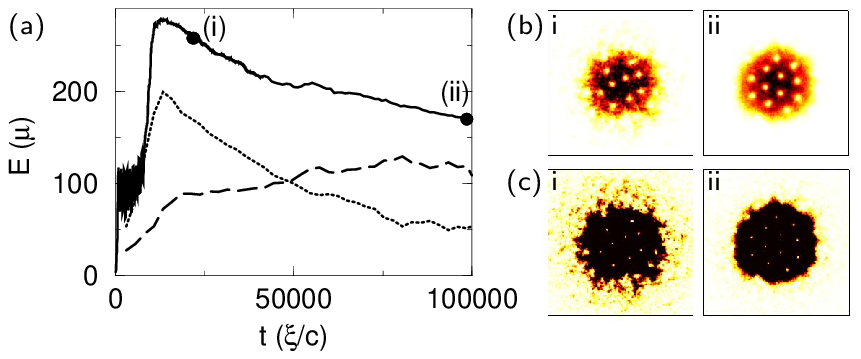}
\caption{Crystallisation: (a) Total condensate energy (solid
line), sound energy (dotted line) and vortex energy (dashed line).
Snapshots of the (a) total condensate density $(0-n_0)$ and (b)
low density $(0-0.05n_0)$ at times (i) $20000$ and (ii)
$100000~(\xi/c)$.}
\end{figure}

In our simulations, the finite grid size sets a maximum value in
momentum space, with higher $k$ modes having zero occupation
\cite{norrie05}. We have imposed reduced values of the momentum
cutoff and consistently observe lattice formation, with no marked
effect on its timescale. This further supports the idea that the
vortex lattice formation is independent of thermal effects.

In the vortex lattice experiments \cite{madison2000,hodby}, the
lattices are observed after up to 1s of trap rotation. In our
simulation, a noisy lattice has formed by $t\sim2{\rm s}$
(Fig.~5(b,i)), while it takes several more seconds for the
vortices to settle into a clean lattice.  However, as shown in
Fig.~3, the timescale of the fragmentation stage is sensitive to
the degree of symmetry-breaking effects present in the system, as
well as the trap rotation frequency and ellipticity.  One would
therefore expect that in a real system, with all its inherent
imperfections, the timescale for this stage will be reduced.

Note that if the rotation is terminated before the peak energy has
been coupled into the system, the lattice still forms, albeit at a
lower energy.  This allows control over the number of vortices
which ultimately form in the lattice.

In summary, we have shown that `stirring' atomic condensates
generates turbulence. We verify that a $k^{-5/3}$ power law is
observed for two-dimensional superfluid turbulence. The turbulent
state subsequently evolves into a vortex lattice by vortex-sound
interactions. Future work will focus on turbulence in spinor
systems and the effect of dimensionality on the turbulence
spectrum.

We acknowledge UK EPSRC for funding and S. A. Gardiner for
discussions.

\end{document}